\documentclass[conference]{IEEEtran}
\usepackage{cite}
\usepackage{url}

\usepackage[dvipsnames,svgnames]{xcolor}
\usepackage{graphicx}
\usepackage{algorithmic}
\usepackage{subfigure}
\usepackage{psfrag}
\DeclareGraphicsExtensions{.xps}
\usepackage{epstopdf} 
\usepackage[cmex10]{amsmath}
\usepackage{amssymb}
\usepackage{amsthm}

\newtheorem{lemma}{Lemma}
\newtheorem{corollary}{Corollary}

\usepackage{eso-pic}

\newcommand{\vect}[1]{\mathbf{#1}}

\def\tr{\mathrm{tr}}

\def\Htran{\mbox{\tiny $\mathrm{H}$}}

\def\CN{\mathcal{N}_{\mathbb{C}}} 
\def\taupu{\tau_\mathrm{p}} 
\def\tauc{\tau_\mathrm{c}} 

\begin{document}

\title{Spectral Efficiency Analysis in \\Dense Massive MIMO Networks\vspace{-.5cm}}
\author{
\IEEEauthorblockN{FahimeSadat Mirhosseini\IEEEauthorrefmark{1}\IEEEauthorrefmark{2}, Andrea Pizzo\IEEEauthorrefmark{2}, Luca Sanguinetti\IEEEauthorrefmark{2}, Aliakbar Tadaion\IEEEauthorrefmark{1}}
\IEEEauthorblockA{\IEEEauthorrefmark{1}\small{Department of Electrical Engineering, Yazd University, Yazd, Iran}}
\IEEEauthorblockA{\IEEEauthorrefmark{2}\small{Dipartimento di Ingegneria dell'Informazione, University of Pisa, Pisa, Italy}\vspace{-.2cm}}

\vspace{-.9cm}}
\maketitle
\begin{abstract}
This work considers the uplink of a Massive MIMO
network wherein the base stations (BSs) are randomly deployed
according to a homogenous Poisson point process of intensity
$\lambda$. Each BS is equipped with $M$ antennas and serves $K$ user
equipments. A rigorous stochastic geometry framework with
a multi-slope path loss model and pilot-based channel estimation
is used to analyze the impact of $\lambda$ on channel estimation
accuracy and spectral efficiency. Both maximum ratio and zero-
forcing combiners are considered. Interesting analytical insights are provided
into the interplay of network parameters such as $\lambda$, antenna-UE ratio $M/K$, and pilot reuse factor. The relative strength
of pilot contamination and (inter- and intra-cell) interference
is analytically and numerically evaluated, as a function of $\lambda$. It
turns out that pilot contamination becomes relevant only for impractical
values of $M/K\ge 100$.
\end{abstract}

\vspace{-.1cm}
\section{Introduction}\label{sec:Introduction}
The data traffic in cellular networks has grown at an exponential pace for decades, thanks to the continuous evolution of the wireless technology. The traditional way to keep up with the traffic growth is to allocate more frequency spectrum.
Consequently, there is little bandwidth left in the sub-6\,GHz bands that are attractive for wide-area coverage. Another way is to reduce cell sizes by deploying more and more base stations (BSs) \cite{AndrewsMagazine}. Quantitatively speaking, next generation of cellular networks foresees three different degrees of BS densification (on the basis of traffic load) \cite{ge20165g}: low dense with roughly a BS density $\lambda \le 10$ (measured in ${\rm BS / km^2}$), dense with $10 < \lambda < 100$ and ultra dense with $\lambda \ge 100$.

Typically, higher BS density leads to an irregular network deployment, which is well described by stochastic geometry tools \cite{Andrews2014a}. These tools were used in \cite{andrews2011tractable} for single-input single-output (SISO) networks to show that the spectral efficiency (SE) is a monotonic non-decreasing function of $\lambda$. This result was proved by using a distance-independent path loss model. A more realistic model accounts for a multi-slope path loss, wherein the path loss exponent depends on the distance between user equipments (UEs) and BSs \cite{zhang2015downlink,pizzo2018network}. With such a model, different operating regimes can be identified (as function of $\lambda$) for which an increase, saturation, or decrease of the SE is observed \cite{ding2017performance}. Despite this, multi-slope path loss models are not frequently used in cellular networks as they make the theoretical analysis much more demanding. An analytical simplification that still captures the essence of the model is discussed in \cite{Kountouris2016}, wherein the power decadence of the radiated signal is divided in two spatial half-spaces.

Nowadays, densification is not the only weapon in the hands of operators to increase the capacity of cellular networks. A very promising technology in this direction is Massive MIMO \cite{marzetta2010noncooperative,Larsson2014}, wherein BSs are equipped with a large number $M$ of low-power, fully digital controlled, and physically small antennas to serve a multitude $K$ of UEs by spatial multiplexing. 
A solid and mature theory for Massive MIMO has been developed in recent years, as underlined by the two recent textbooks \cite{Marzetta2016a,massivemimobook}. 
Remarkably, the long-standing pilot contamination issue, which was believed to impose a fundamental limitation \cite{marzetta2010noncooperative}, has recently been resolved in spatially correlated channels by using multicell processing \cite{BHS18A,SanguinettiBH19}.

The aim of this work is to investigate the impact of BS
densification on the uplink (UL) SE of Massive MIMO networks. The analysis is
conducted on the basis of the theoretical framework developed
in \cite{pizzo2018network}, which provides SE lower bounds in closed form by
using a rigorous stochastic geometry-based framework and a
general multi-slope path loss model. Analytical and numerical
results are given for both maximum-ratio (MR) and zero-
forcing (ZF) combining schemes by using the simplified dual-slope path loss model in \cite{Kountouris2016}. This allows to
gain several insights into the interplay among different network
parameters, e.g., BS density, antenna-UE ratio $M/K$, and
pilot reuse factor. Interesting observations are also made with
respect to the impact and relative importance of (intra- and
inter-) interference and pilot contamination in dense networks.

\section{Network Model}\label{sec:Problem_Statement}

We consider the UL of a Massive MIMO cellular network wherein the BSs are spatially distributed at locations ${\{\mathbf{x}_l\}\subset \mathbb{R}^2}$ according to a homogenous Poisson point process ${\Psi_{\lambda} = \{\mathbf{x}_l; \, l \in \mathbb{N} \}}$ of intensity ${\lambda}$. 
Each BS has $M$ antennas and serves simultaneously $K$ single-antenna UEs. 
We use the nearest BS association rule so that the coverage area of a BS is its Poisson-Voronoi cell wherein the $K$ UEs are uniformly distributed. 
The network operates according to the classical synchronous time-division-duplex protocol (TDD) over each coherence block \cite{massivemimobook}, which is composed of $\tauc$ complex-valued samples.
In each coherence block, $\taupu = K \zeta$ samples with $\zeta$ being the pilot reuse factor, are used for acquiring channel state information by means of UL pilot sequences.

\subsection{Channel model}
The channel $\vect{h}_{li}^{j} \in \mathbb{C}^{M}$ between the UE~$i$ in cell~$l$ and BS~$j$ is modeled as i.i.d. Rayleigh fading:
\begin{align}\label{eq:iid-Rayleigh fading}
\vect{h}_{li}^{j} \sim \CN( \vect{0}_M, \beta_{li}^{j} \vect{I}_M)
\end{align}
where $\beta_{li}^{j}$ is computed on the basis of a multi-slope path loss model\cite{zhang2015downlink}
\begin{eqnarray}\label{pathloss_model}
\beta_{li}^{j} = \Upsilon_n (d_{li}^{j})^{-\alpha_n}
\end{eqnarray}
where ${d_{li}^{j} \in [R_{n-1}, R_{n})}$, for ${n = 1, \ldots, N}$ denotes the distance of UE $i$ in cell $l$ from BS $j$, ${0 \leq \alpha_1 \leq \cdots \leq \alpha_N}$ are the power decay factors, ${0 = R_0 < \cdots < R_N = \infty}$ denote the distances at which a change in the power decadence occurs, whereas ${\Upsilon_{n+1} = \Upsilon_n R_{n}^{\alpha_{n+1} - \alpha_{n}}}$ for $n = 1, \ldots, N-1$ are chosen for continuity purposes of the model with $\Upsilon_1$ being a design parameter.
Clearly, setting $N = 1$ yields the widely used single-slope path loss model, i.e., ${\beta_{li}^{j} = \Upsilon_1 (d_{li}^{j})^{-\alpha_1}}$.

\begin{figure}[t!]
 \centering
 \includegraphics[width=0.52\textwidth]{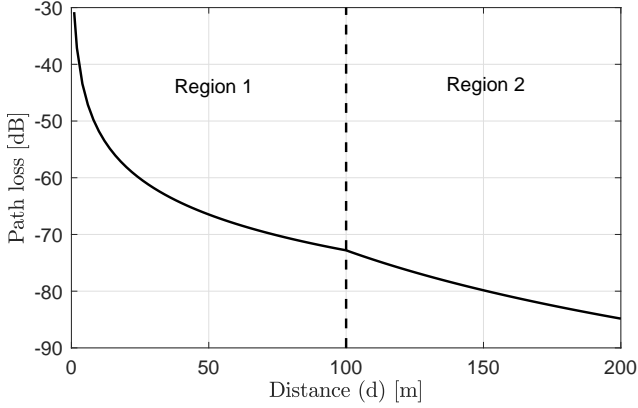} 
 \caption{Multi-slope path loss [dB] for the parameters listed in Table~\ref{table1}.}
 \label{fig1}
 \end{figure}

Although the subsequent analysis is valid for any path loss model, we follow \cite{AndrewsMagazine, zhang2015downlink, Kountouris2016} and use the dual-slope model, i.e., $N=2$, illustrated in Fig.~\ref{fig1}, with parameters reported in Table~\ref{table1}. Particularly, we choose ${R_1= 100}$ m\footnote{Typical values for ITU-R UMi model range from $20$ to $200$ m \cite{AndrewsMagazine}.} and assume $\alpha_1 = 2.1$ and $\alpha_2 = 4$. 
The coefficient $\Upsilon_2$ ensures the  continuity between the two regions. As discussed in \cite{AndrewsMagazine, zhang2015downlink, Kountouris2016}, this model can be seen as a simplified version of the ITU-R UMi model \cite{3GPP}.

%

\begin{table}[t!]
\centering
\caption{Two-slope path loss model.} 
\begin{tabular}{l|l|l}
   $R_n$ & $\Upsilon_n$ & $\alpha_n$ \\
\hline

$R_1=100$ & $\Upsilon_1= 8.3e-04 $, $\Upsilon_2= 5.2481$& $\alpha_1=2.1$, $\alpha_2=4$ 
\end{tabular}\label{table1}
\end{table}
\subsection{Pilot allocation and power control}
We assume that a pilot book $\{{\boldsymbol{\Phi}} \in \mathbb{C}^{\taupu \times \taupu}\}$ of $\taupu$ orthonormal UL pilot sequences ${\{\boldsymbol{\phi}_{li}\}_{i=1}^K}$ is used for channel estimation in each cell $l \in \Psi_{\lambda}$.
To avoid cumbersome pilot coordination as the network densifies, we assume that in each coherence block the BS $l$ picks uniformly
at random  a subset of $K$ different sequences from $\boldsymbol{\Phi}$ and distributes them among its served UEs \cite{bjornson2016deploying}. Since
${\tau_p= \zeta K}$, we have that the reuse factor is ${\zeta = \tau_p / K > 1}$.
In other words, there are on average $\mathbb{E}\{\Psi_\lambda\} / \zeta$ cells that share the same pilot subset. This is modeled in
each cell through a Bernoulli stochastic variable ${a_{l' l} \sim \mathcal{B}({1/\zeta})}$
for ${l' \neq l}$ and ${a_{l l} = 1}$. 
 \begin{figure}[t]
 \centering
   { \includegraphics[width=0.52\textwidth]{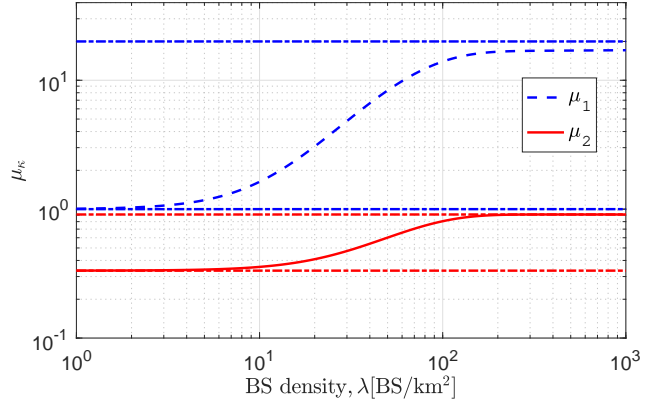}\label{fig2}}
 \caption{Behavior   of $\mu_1$ and $\mu_2$ as a function of $\lambda$ $[\rm{BS/km^2}]$ for the path loss model in Table~\ref{table1}.}\label{fig2}\vspace{-0.5cm}
 \end{figure}
Following \cite{bjornson2016deploying}, we assume that each UE uses a statistical channel inversion power-control policy such that
\begin{equation} \label{ULpower}
p_{li} = \rho/{\beta_{li}^{l}}
\end{equation}
where $\rho$ is a design parameter. This ensures a uniform signal-to-noise ratio ${\mathsf{SNR}_0 = \rho/\sigma^2}$ across the network \cite{bjornson2016deploying}.

 \begin{figure*}
\begin{eqnarray}\label{mu_kappa}
\mu_\kappa = 2 \sum_{n = 1}^{N} \frac{\Gamma\left(2; \pi \lambda R_{n-1}^2\right) - \Gamma\left(2; \pi \lambda R_{n}^2\right)}{\kappa \alpha_n - 2} + \frac{2 c_n(\kappa)}{(\pi \lambda)^{\frac{\kappa \alpha_n}{2} -1}} 
\left(  \Gamma\left(1+\frac{\kappa \alpha_n}{2}; \pi \lambda R_{n-1}^2\right) -  \Gamma\left(1+\frac{\kappa \alpha_n}{2}; \pi \lambda R_{n}^2\right)\right), \; \kappa=1,2
\end{eqnarray}
\hrule
\end{figure*}


 \section{Network Analysis}\label{sec:net_analyz}
The aim is to analyze the channel estimation accuracy and SE of the investigated Massive MIMO network as the BS density $\lambda$ increases. 
The impact and relative importance of pilot contamination, induced by the so-called coherent interference \cite{massivemimobook}, is also investigated.
The analysis is conducted for the ``typical UE'', which is statistically representative for any other UE in the network \cite{Baccelli2008a}. Without loss of generality, we assume that the ``typical UE'' has an arbitrary index $k$ and is connected to an arbitrary BS $j$. 

\subsection{Preliminaries}
For later convenience, let us define the average coefficients $\mu_\kappa$ with $\kappa=1,2$ as in \eqref{mu_kappa} where we use\cite{pizzo2018network}
\begin{equation} \label{cn_kappa}
 c_n(\kappa) \!=\! - \frac{R_n^{2 - \kappa \alpha_n}}{ \kappa \alpha_n - 2} \sum_{i = n + 1}^N \left(\frac{\Upsilon_i}{\Upsilon_n}\right)^\kappa \frac{R_{i-1}^{2 - \kappa \alpha_i} - R_i^{2 - \kappa \alpha_i}}{\kappa \alpha_i - 2}.
\end{equation}
Both are needed for the analysis of channel estimation accuracy and SE. If a single-slope model with generic path loss coefficient $\alpha$ is adopted, $\mu_1$ and $\mu_2$ become
\begin{align}
\mu_1 =  \frac{2}{\alpha-2}\\
\mu_2 =  \frac{1}{\alpha-1}
\end{align}
and are thus independent of the BS density $\lambda$.
For the dual-slope model in Table~\ref{table1}, the behavior of $\mu_1$ and $\mu_2$ as a function of $\lambda$ is illustrated in Fig.~\ref{fig2}.
As seen, both coefficients are monotonic non-decreasing functions of $\lambda$ and take values in the following intervals:
\begin{align}\label{mu_kappa_singlelslope_1}
\frac{2}{\alpha_2-2} \le \mu_1 \le \frac{2}{\alpha_1-2}\\ \frac{1}{\alpha_2-1} \le \mu_2 \le \frac{1}{\alpha_1-1}\label{mu_kappa_singlelslope_2}
\end{align}
with $\mu_2$ being much smaller than $ \mu_1$. Fig.~\ref{fig2} shows that the lower bounds are achieved for ${\lambda \le 10}$, which corresponds to a low dense network wherein the probability for a UE to be in the second region of Fig. \ref{fig1} is relatively high. The upper bounds are achieved for an ultra dense network with ${\lambda \ge 10^2}$. In this case, the UEs operate mostly in the first region. For ${10 \le \lambda \le 10^2}$, $\mu_1$ and $\mu_2$ increase monotonically and this basically accounts for scenarios that are a mixture of the two propagation conditions.
Notice that the same trend is observed with a general $N$-slope path loss model. In this case, however, there are $N$ ranges of values of $\lambda$ wherein the coefficients $\mu_1$ and $\mu_2$ are equal to ${\mu_1 =\frac{2}{\alpha_n-2}}$ and ${\mu_2 =\frac{1}{\alpha_n-1}}$ for $n=1,\ldots,N$. 


\subsection{Channel estimation}
We call $\mathbf{Y}^{j} \in \mathbb{C}^{M \times \tau_p}$ the signal received at BS $j$ during UL pilot transmission. The vector $\mathbf{y}_{jk}^{j} = \mathbf{Y}^{j}  \boldsymbol{\phi}_{jk}^{*}$ takes the form: 
\begin{eqnarray}
\mathbf{y}_{jk}^{j} =  \sqrt{\frac{\rho_{\tr}}{\beta_{jk}^{j}}}  \mathbf{h}_{jk}^{j} + \sum_{l\in {\Psi}_{\lambda} \setminus {\{j\}} } { a_{l j} \sqrt{\frac{\rho_{\tr}}{\beta_{li}^{l}}}  \mathbf{h}_{lk}^{j}} + \mathbf{n}_{jkj}
\end{eqnarray}
where $\mathbf{n}_{jkj}^\mathrm{p} \sim \mathcal{N}_{\mathbb{C}} (\mathbf{0}, \sigma^2 \mathbf{I}_M)$ and ${\rho_{\tr}}$ is a design parameter.
The minimum mean-squared error (MMSE)
estimate of $\mathbf{h}_{jk}^j$ based on $\mathbf{y}_{jk}^{j}$ is \cite[Sec. 3.2 ]{massivemimobook}
\begin{eqnarray}\label{channelEst}
\widehat{\mathbf{h}}_{jk}^j = \gamma_{jk}^{j} \frac{ 1}{\sqrt{\beta_{jk}^{j} \tau_p \rho_{\tr}}} \mathbf{y}_{jk}^{j}\sim \mathcal{N}_{\mathbb{C}}(\mathbf{0}, \gamma_{jk}^{j} \mathbf{I}_M)
\end{eqnarray}
with
\begin{equation}\label{eq:sec4_delta_1}
\gamma_{jk}^{j} =  \beta_{jk}^{j}  \left(1 + \sum\limits_{l \in {\Psi}_\lambda \setminus \{j\} }  a_{lj}     \frac{\beta_{l k}^{j}}{\beta_{l k}^{l}} +  \frac{1}{\tau_p}\frac{1}{\mathsf{SNR}_0}\right)^{-1}.
\end{equation} 
The average NMSE is given by
\begin{align} \notag 
\mathsf{NMSE}  & \triangleq \mathbb{E}_{\{\bf d, a\}} \left\{\frac{ \mathbb{E}_{{\{\bf h\}}} \{\| \vect{h}_{jk}^{j} - \widehat{\vect{h}}_{jk}^{j} \|^2\}}{ \mathbb{E}_{{\{\bf h\}}}\{\| \vect{h}_{jk}^{j} \|^2\}}\right\}   \\& \label{NMSE_Uncorr}
= 1 - \mathbb{E}_{{\{\bf d\}}} \left\{\frac{1}{\beta_{jk}^{j}}\mathbb{E}_{{\{\bf a\}}} \left\{\gamma_{jk}^{j}\right\}\right\}
\end{align}
as it follows by taking the expectation with respect to the channel distributions. The following result is thus obtained.  
\begin{figure}[t!]\vspace{-0.5cm}
\centering
\includegraphics[width=0.52\textwidth]{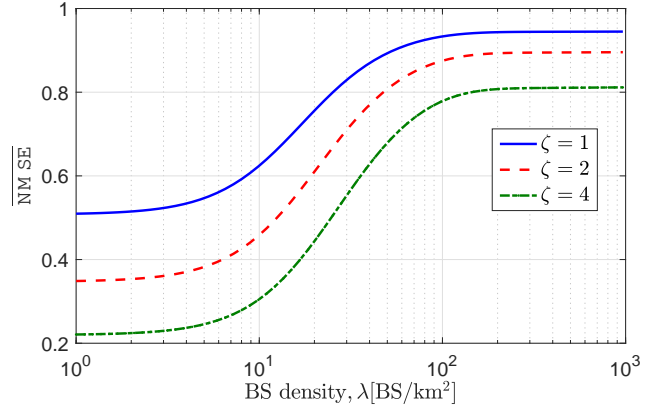}
 \caption{$\overline{\mathsf{NMSE}}$ as a function of $\lambda$ $[\rm{BS/km^2}]$ for the path loss model in Table~\ref{table1} for $\zeta = \{1,2,4\}$ and $\mathsf{SNR}_0 = 5$ dB.}\label{fig:NMSE}\vspace{-0.5cm}
\end{figure}

\begin{corollary}[Average NMSE] \label{corollary_NMSE} 
The average NMSE can be upper bounded as:
\begin{align}\label{NMSE}
\mathsf{NMSE} \le \overline{\mathsf{NMSE}} = 1 - 1/A(\mu_1)
\end{align}
with
\begin{align}\label{A_definition}
A(\mu_1) = 1 + \frac{\mu_1}{\zeta} + \frac{1}{{\zeta} K}\frac{1}{ \mathsf{SNR}_0}.
\end{align}
\end{corollary} 
\begin{IEEEproof}
The proof follows from \cite[Appendix~A]{pizzo2018network}.
\end{IEEEproof}

\begin{figure*}
\begin{align}\label{snr_MR}
\gamma^{\rm MR} &= \frac{1}{\underbrace{\frac{A(\mu_1)}{M \mathsf{SNR}_0}}_{\textnormal{Noise}} + \underbrace{ \frac{K}{M}A(\mu_1)}_{\textnormal{ Intra-cell \,Interference}} + \underbrace{\frac{K}{M}\left(A(\mu_1)\mu_1 + \frac{\mu_2}{\zeta}\right)}_{\textnormal{Inter-cell \, Interference}} + \underbrace{\frac{\mu_2}{\zeta}}_{\textnormal{Pilot\, Contamination}}}\\\label{snr_ZF}
\bigskip
\gamma^{\rm ZF} &= \frac{1}{\underbrace{\frac{A(\mu_1)}{(M-K) \mathsf{SNR}_0}}_{\textnormal{Noise}} + \underbrace{ \frac{K}{M-K}(A(\mu_1)-1)}_{\textnormal{Intra-cell \,Interference}} + \underbrace{\frac{K}{M-K}A(\mu_1)\mu_1}_{\textnormal{Inter-cell \, Interference}} +\underbrace{\frac{\mu_2}{\zeta}}_{\textnormal{Pilot\, Contamination}}}
\end{align}
\hrule\vspace{-0.2cm}
\end{figure*}

\smallskip
Corollary~\ref{corollary_NMSE} shows that on average the NMSE is affected by pilot-sharing UEs through $ {\mu_1}/{\zeta}$.
Therefore, it reduces as the pilot reuse factor $\zeta$ decreases and/or the BS density $\lambda$ increases, which is due to the monotonicity of $\mu_1$ with respect to $\lambda$.
This is exemplified in Fig.~\ref{fig:NMSE} where $\overline{\mathsf{NMSE}}$ is plotted as a function $\lambda$ for $\zeta \in \{1,2,4\}$ and $\mathsf{SNR}_0 = 5$ dB.
As seen, $\overline{\mathsf{NMSE}}$ is lower and upper bounded and takes values in the interval:
\begin{align}\label{NMSE_singlelslope}
1 - \frac{1}{A\left(\frac{2}{\alpha_2-2}\right)}  \le \overline{\mathsf{NMSE}} \le 1 - \frac{1}{A\left(\frac{2}{\alpha_1-2}\right)}
\end{align}
where $A\big(\frac{2}{\alpha_2-2}\big)$ and $A\big(\frac{2}{\alpha_1-2}\big)$ are obtained after substituting the limits of $\mu_1$ given by \eqref{mu_kappa_singlelslope_1} into \eqref{A_definition}. The $\overline{\mathsf{NMSE}}$ is less than $0.6$ for any $\zeta$ when ${\lambda \le 10}$, i.e., a low dense network, but it increases fast as $\lambda$ grows. In an ultra dense regime with ${\lambda \ge 10^2}$, e.g., it is higher than $0.8$. Particularly, it approaches $0.95$ with $\zeta=1$. As shown later, this will have a strong negative impact on the SE. We notice also that increasing $\zeta$ brings less benefits (in terms of estimation accuracy) as $\lambda$ becomes large. This is because the cell size decreases with $\lambda$ and thus reducing the interference of pilot-sharing UEs by increasing their average distance has little impact as the network becomes denser and denser.  
Interestingly, $\overline{\mathsf{NMSE}}$ is very sensitive to densification in the interval ${10 \le \lambda \le 10^2}$. This implies that any additional BS in this regime comes with a higher cost (in terms of channel estimation accuracy) with respect to both cases ${\lambda \le 10}$ and ${\lambda \ge 10^2}$.

\subsection{Spectral Efficiency} 
We denote by ${{\bf v}_{jk} \in \mathbb {C}^{M}}$ the receive combining vector associated with UE $k$ in cell $j$. Two popular choices for $\vect{v}_{jk}$ are MR and ZF \cite{Marzetta2016a}:
\begin{equation} \label{combining_schemes}
\vect{V}_{j} \triangleq \left[ \vect{v}_{j 1} \, \ldots \, \vect{v}_{j K}  \right] = \begin{cases}
 \widehat{\vect{H}}_{j}^{j} & \!\!\!\textrm{with MR} \\ 
\widehat{\vect{H}}_{j}^{j} \left(
 (  \widehat{\vect{H}}_{j}^{j})^{\Htran} \widehat{\vect{H}}_{j}^{j}  \right)^{-1} & \!\!\!\textrm{with ZF}
\end{cases}
\end{equation}
with ${\widehat{\vect{H}}_{j}^{j}  = [\widehat{\vect{h}}_{j 1}^{j} \, \ldots \, \widehat{\vect{h}}_{j K}^{j}]\in \mathbb{C}^{M \times K}}$ containing the estimates of intra-cell channels in cell $j$. In a multicell Massive MIMO network with imperfect knowledge of the channel, they are both suboptimal \cite{massivemimobook}, but widely applied in the literature because of their analytical tractability through the use-and-then-forget (UatF) bound for SE\cite{Marzetta2016a}. For the investigated network, the following result is obtained.
\begin{lemma}[Average ergodic SE \cite{pizzo2018network}]\label{theorem_SE}
If MMSE channel estimates are used and the UL powers are chosen as in \eqref{ULpower}, a lower bound on the UL average ergodic channel capacity is:
\begin{align}\label{lower_bound_SE}
\underline{{\mathsf{SE}}} =  \left(1 - \frac{K \zeta}{\tau_c}\right) \log_2(1+\gamma) 
\end{align}
where $\gamma$ is given by \eqref{snr_MR} and \eqref{snr_ZF} for MR and ZF, respectively. 
\end{lemma}
 One can make interesting observations from Lemma~\ref{theorem_SE}.
With both schemes, the interference is decomposed into intra-cell and inter-cell interference. The first one accounts for the interference generated by the UEs located in the Poisson-Voronoi region of the serving BS, while the second one is due to all the other UEs. As seen, both terms reduce linearly with $M$ if MR is used while they reduce with ${M - K}$ when ZF is employed.\footnote{This is why the sum of the two is generally referred to as non-coherent interference \cite{massivemimobook}.}
This is because ZF sacrifices $K$ spatial dimensions to suppress intra-cell interference. The same happens for the noise contribution. Moreover, we notice that both noise and (intra- and inter-) interference depend also on ${A(\mu_1)}$, which is due to the imperfect knowledge of the channel; see Corollary~\ref{corollary_NMSE}. Since ${A(\mu_1)}$ increases linearly with $\mu_1$, both noise and interference grow as $\lambda$ increases. The inter-cell interference term increases at an even faster rate with $\lambda$ since it is a function of ${A(\mu_1)}\mu_1$. 
%
%
The last term $\mu_2/\zeta$ in \eqref{snr_MR} and \eqref{snr_ZF} accounts for pilot contamination since it is independent of the antenna-UE ratio $M/K$\cite{massivemimobook}. Interestingly, it is the same for both schemes. The following corollary is easily obtained.
\begin{corollary}\label{corollary_2}
When ${M\to \infty}$, ${\gamma^{\rm ZF} =\gamma^{\rm MR} \asymp \gamma_{M\to \infty}}$ with $\gamma_{M\to \infty} = {\zeta}/{\mu_2}$
and the ultimately achievable rate is 
\begin{align}\label{R_inf}
R_{M\to \infty} = \left(1 - \frac{\zeta K}{\tau_c}\right)\log_2\left(1 + \frac{\zeta}{\mu_2}\right).
\end{align}
\end{corollary} 

\begin{figure}[t!]\vspace{-0.5cm}
\centering
\includegraphics[width=0.52\textwidth]{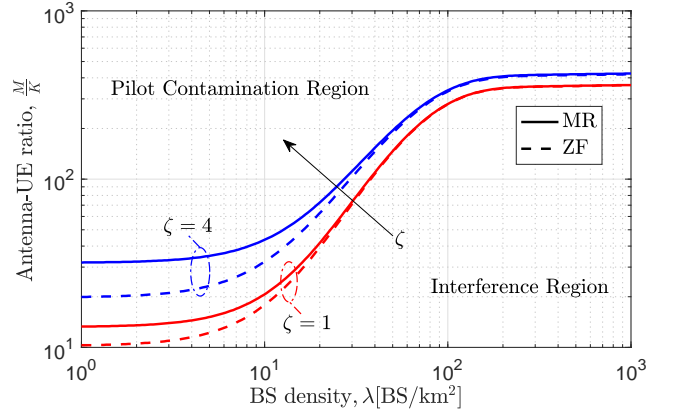}
 \caption{Antenna-UE ratio $M/K$ needed for the interference to equal pilot contamination as a function of $\lambda$. The dual-slope path loss model in Table \ref{table1} is adopted. Also, $\zeta = \{1,4\}$ 
 and $K=10$.}
\label{fig:Region}\vspace{-0.5cm}
\end{figure}

As shown in Fig.~\ref{fig2}, $\mu_2$ is much smaller than $ \mu_1$ for any $\lambda$. This implies that interference in \eqref{snr_MR} and \eqref{snr_ZF} is likely the dominating term of $\gamma$.
To provide evidence of this, Fig.~\ref{fig:Region} illustrates the antenna-UE ratio $M/K$ for which the interference and pilot contamination in \eqref{snr_MR} and \eqref{snr_ZF} are equal to each other.
The curves are plotted as a function of $\lambda$ for ${\zeta\in \{1,4\}}$ and must be understood in the sense that in the region above each one pilot contamination is higher than interference. 
As expected, for any $\lambda$ the pilot contamination region with MR is smaller than that with ZF. This is because unlike MR, ZF mitigates the effect of intra-cell interference. The gap decreases as $\lambda$ increases since inter-cell interference becomes more relevant. 
Remarkably, we observe that for practical values of $M/K$ in the range interval ${4\le M/K \le 10}$\cite{massivemimobook}, we never fall into the pilot contamination region regardless of $\lambda$ and $\zeta$. If large values are considered, i.e., $10 \le M/K \le 100$, interference always
dominates pilot contamination for $\lambda \ge 50$. More generally, the
results of Fig.~\ref{fig:Region} show that in dense and ultra-dense Massive
MIMO networks interference is the major impairment. Pilot
contamination becomes relevant only for impractical values of
$M/K$.

\begin{figure}[t]
\centering
\includegraphics[width=0.52\textwidth]{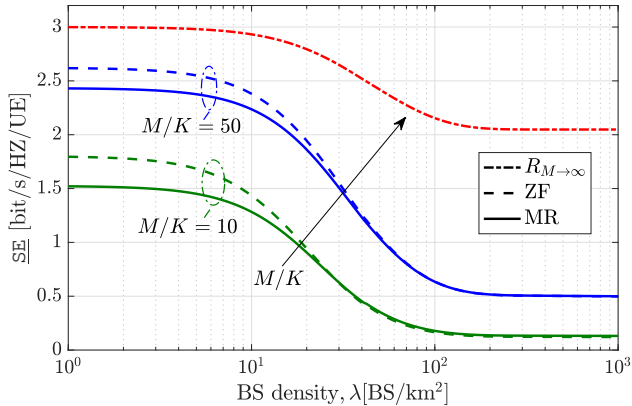}
\caption{$\underline{{\mathsf{SE}}}$ as a function of $\lambda$ $[\rm{BS/km^2}]$ for the dual-slope path loss model in Table \ref{table1} with $M/K = \{10, 50\}$, $K = 10$ and $\mathsf{SNR}_0 = 5$ dB.}\label{fig:SE}\vspace{-0.5cm}
\end{figure}

To quantify the UL SE, Fig.~\ref{fig:SE} plots the lower bound in \eqref{lower_bound_SE} versus $\lambda$ with MR and ZF. Both schemes use the optimal pilot reuse factor, obtained by exhaustive search. We consider the practical value ${M/K=10}$ but also ${M/K=50}$. This latter case is used to evaluate the gap with the ultimately achievable rate in \eqref{R_inf}, which is also reported in Fig.~\ref{fig:SE}. The asymptotically optimal pilot reuse factor is computed from Corollary \ref{corollary_2} as follows. 
\begin{corollary}\label{corollary_3} 
When ${M\to \infty}$, the optimal pilot reuse factor $\zeta$ that maximizes \eqref{R_inf} is 
\begin{align}\label{zeta_inf}
\zeta_{M\to \infty}^{\mathsf{opt}} = \mu_2 \left( \frac{\nu}{W(\nu e)} - 1\right)
\end{align}
where $W(.)$ is Lambert function and $\nu = 1 + {\tau_c}/({\mu_2 K})$.
\end{corollary} 
\begin{IEEEproof}
By taking the derivative of \eqref{R_inf} yields
\begin{eqnarray}
\frac{\partial R_{M\to \infty}}{\partial \zeta} = -\frac{K}{\tau_c} \log_2\left(1 + \frac{\zeta}{\mu_2}\right) + \frac{1 - \zeta K / \tau_c}{\mu_2 \ln(2) (1 + \zeta/ \mu_2)}.
\end{eqnarray}
Setting $ x = \frac{\tau_c/ K - \zeta }{\mu_2 + \zeta}$ we obtain
$ (x + 1) e^{x+1} = e (\frac{\tau_c}{K \mu_2} + 1)$, whose solution is easily found as $ x = W( e ( \frac{\tau_c}{K \mu_2} + 1)) - 1$.
\end{IEEEproof}
\smallskip
From Fig.~\ref{fig:SE}, it is seen that $\underline{{\mathsf{SE}}}$ is a monotonic non-increasing function of $\lambda$. This is because both interference and pilot contamination increase with $\lambda$ through $\mu_1$ and $\mu_2$. 
Although ZF outperforms MR for $\lambda \le 10$, the gain reduces as $\lambda$ increases. Both schemes achieve the same performance for $\lambda > 30$. 
This is because inter-cell interference becomes the dominating term in \eqref{snr_MR} and \eqref{snr_ZF} when the BSs are much closer to each other. We also notice that the gap with the ultimate achievable rate is substantial and increases with $\lambda$. Particularly, in an ultra dense network with ${\lambda = 50}$, MR and ZF achieve roughly ${15\%}$ of $R_{M\to \infty}$ with ${M/K = 10}$. In the extreme case of ${M/K = 50}$, the gap is still large, though ${45\%}$ of $R_{M\to \infty}$ is achieved.
%

Finally, we observe that the SE per UE maintains approximately constant in low dense, i.e., $\lambda \le 10$, and ultra dense, i.e., $\lambda \ge 10^2$, networks.
However, the area SE in $\rm{bit/s/Hz/km^2}$ given by
\begin{equation}
\underline{\mathsf{ASE}} = \lambda K \, \underline{{\mathsf{SE}}}
\end{equation} 
increases linearly with $\lambda$ in these regions. This implies that there are operating regimes in which we can linearly increase the ASE while keeping the same SE per UE across the network. {While for $10 < \lambda < 10^2$, ASE will suffer from a small decrease.}


\section{Conclusions}\label{sec:Conclusion}

We studied the impact of BS densification on channel estimation and SE in the UL of Massive MIMO. The analysis was conducted for MR and ZF with a two-slope path loss model. Lower bounds of the NMSE and SE were provided in closed forms from which interesting insights were obtained. Particularly, we quantified the relative importance of pilot contamination and (intra- and inter-) interference, and showed that the latter largely dominates for any BS density $\lambda$
irrespective of the antenna-UE ratio $M/K$. We showed that the SE is a monotonic non-increasing function of $\lambda$, and also demonstrated that both MR and ZF provide low SE already for ${\lambda \ge 10}$ $[\rm{BS/km^2}]$, even when $M/K$ takes large values (up to $50$). This means that multicell signal processing (e.g., \cite{BHS18A,SanguinettiBH19}) is required if a dense Massive MIMO network needs to be deployed. \vspace{-0.2cm}

\bibliographystyle{IEEEtran}
\bibliography{IEEEabrv,ref,ref_book,ref_1}

\begin{thebibliography}{10}
\providecommand{\url}[1]{#1}
\csname url@samestyle\endcsname
\providecommand{\newblock}{\relax}
\providecommand{\bibinfo}[2]{#2}
\providecommand{\BIBentrySTDinterwordspacing}{\spaceskip=0pt\relax}
\providecommand{\BIBentryALTinterwordstretchfactor}{4}
\providecommand{\BIBentryALTinterwordspacing}{\spaceskip=\fontdimen2\font plus
\BIBentryALTinterwordstretchfactor\fontdimen3\font minus
  \fontdimen4\font\relax}
\providecommand{\BIBforeignlanguage}[2]{{%
\expandafter\ifx\csname l@#1\endcsname\relax
\typeout{** WARNING: IEEEtran.bst: No hyphenation pattern has been}%
\typeout{** loaded for the language `#1'. Using the pattern for}%
\typeout{** the default language instead.}%
\else
\language=\csname l@#1\endcsname
\fi
#2}}
\providecommand{\BIBdecl}{\relax}
\BIBdecl

\bibitem{AndrewsMagazine}
J.~G. {Andrews}, X.~{Zhang}, G.~D. {Durgin}, and A.~K. {Gupta}, ``Are we
  approaching the fundamental limits of wireless network densification?''
  \emph{IEEE Communications Magazine}, vol.~54, no.~10, pp. 184--190, October
  2016.

\bibitem{ge20165g}
X.~Ge, S.~Tu, G.~Mao, C.-X. Wang, and T.~Han, ``{5G} ultra-dense cellular
  networks,'' \emph{IEEE Wireless Communications}, vol.~23, no.~1, pp. 72--79,
  2016.

\bibitem{Andrews2014a}
J.~G. Andrews, S.~Buzzi, W.~Choi, S.~V. Hanly, A.~Lozano, A.~C.~K. Soong, and
  J.~C. Zhang, ``What will {5G} be?'' \emph{{IEEE} J. Sel. Areas Commun.},
  vol.~32, no.~6, pp. 1065--1082, 2014.

\bibitem{andrews2011tractable}
J.~G. Andrews, F.~Baccelli, and R.~K. Ganti, ``A tractable approach to coverage
  and rate in cellular networks,'' \emph{IEEE Transactions on Communications},
  vol.~59, no.~11, pp. 3122--3134, 2011.

\bibitem{zhang2015downlink}
X.~Zhang and J.~G. Andrews, ``Downlink cellular network analysis with
  multi-slope path loss models.'' \emph{IEEE Trans. Communications}, vol.~63,
  no.~5, pp. 1881--1894, 2015.

\bibitem{pizzo2018network}
A.~Pizzo, D.~Verenzuela, L.~Sanguinetti, and E.~Bj{\"o}rnson, ``Network
  deployment for maximal energy efficiency in uplink with multislope path
  loss,'' \emph{IEEE Transactions on Green Communications and Networking},
  2018.

\bibitem{ding2017performance}
M.~Ding and D.~L{\'o}pez-P{\'e}rez, ``Performance impact of base station
  antenna heights in dense cellular networks,'' \emph{IEEE Transactions on
  Wireless Communications}, vol.~16, no.~12, pp. 8147--8161, 2017.

\bibitem{Kountouris2016}
J.~Arnau, I.~Atzeni, and M.~Kountouris, ``Impact of los/nlos propagation and
  path loss in ultra-dense cellular networks,'' in \emph{2016 IEEE
  International Conference on Communications (ICC)}, May 2016, pp. 1--6.

\bibitem{marzetta2010noncooperative}
T.~L. Marzetta, ``Noncooperative cellular wireless with unlimited numbers of
  base station antennas,'' \emph{IEEE Transactions on Wireless Communications},
  vol.~9, no.~11, pp. 3590--3600, 2010.

\bibitem{Larsson2014}
E.~G. Larsson, F.~Tufvesson, O.~Edfors, and T.~L. Marzetta, ``Massive {MIMO}
  for next generation wireless systems,'' \emph{IEEE Commun. Magazine},
  vol.~52, no.~2, pp. 186--195, Feb. 2014.

\bibitem{Marzetta2016a}
T.~L. Marzetta, E.~G. Larsson, H.~Yang, and H.~Q. Ngo, \emph{Fundamentals of
  {M}assive {MIMO}}.\hskip 1em plus 0.5em minus 0.4em\relax Cambridge
  University Press, 2016.

\bibitem{massivemimobook}
E.~Bj\"{o}rnson, J.~Hoydis, and L.~Sanguinetti, ``Massive {MIMO} networks:
  {Spectral}, energy, and hardware efficiency,'' \emph{Foundations and
  Trends{\textregistered} in Signal Processing}, vol.~11, no. 3-4, pp.
  154--655, 2017.

\bibitem{BHS18A}
E.~Bj{\"o}rnson, J.~Hoydis, and L.~Sanguinetti, ``Massive {MIMO} has unlimited
  capacity,'' \emph{IEEE Transactions on Wireless Communications}, vol.~17,
  no.~1, pp. 574--590, 2018.

\bibitem{SanguinettiBH19}
L.~Sanguinetti, E.~Bj{\"o}rnson, and J.~Hoydis, ``Towards {Massive MIMO 2.0}:
  {Understanding} spatial correlation, interference suppression, and pilot
  contamination,'' \emph{CoRR}, vol. abs/1904.03406, 2019.

\bibitem{3GPP}
3GPP, ``Technical specification group radio access network; evolved universal
  terrestrial radio access (e-utra); further advancements for e-utra physical
  layer aspects (release 9). tr 36.814,'' Tech. Rep., 2010.

\bibitem{bjornson2016deploying}
E.~Bj{\"o}rnson, L.~Sanguinetti, and M.~Kountouris, ``Deploying dense networks
  for maximal energy efficiency: Small cells meet massive mimo,'' \emph{IEEE
  Journal on Selected Areas in Communications}, vol.~34, no.~4, pp. 832--847,
  2016.

\bibitem{Baccelli2008a}
F.~Baccelli and B.~Blaszczyszyn, ``{Stochastic geometry and wireless networks:
  Volume I Theory},'' \emph{Foundations and Trends in Networking}, vol.~3, no.
  3-4, pp. 249--449, 2008.

\end{thebibliography}



\end{document}